\documentclass[doublecol]{epl2} 

\title{Intrinsic Localized Modes in a Nonlinear Electrical Lattice with Saturable Nonlinearity}
\shorttitle{ILM in a Nonlinear Electrical Lattice with Saturable Nonlinearity} 

\author{W. Shi\inst{1} \and S. Shige\inst{1} \and Y. Soga\inst{1} \and M. Sato\inst{1} \and A. J. Sievers\inst{2}}
\shortauthor{W. Shi \etal}

\institute{                    
  \inst{1} Graduate School of Natural Science and Technology, Kanazawa University - Kanazawa, Ishikawa 920-1192, Japan
\\
  \inst{2}Laboratory of Atomic and Solid State Physics, Cornell University - Ithaca, NY 14853-2501, USA
}
\pacs{05.45.-a}{Nonlinear dynamics}
\pacs{63.20.Pw}{Localized modes}
\pacs{05.45.Yv}{nonlinear dynamics of solitons}

\abstract{
This experimental study of driven intrinsic localized modes (ILMs) in an electronic circuit lattice with saturable nonlinearity follows the theoretical work of Had\v zievski and coworkers. They proposed that a saturable nonlinearity could introduce transition points where localized excitations in nonintegrable lattices would move freely. In our experiments MOS capacitors provide the saturable nonlinearity in an electric lattice. Because of the soft nonlinearity driver locked, auto-resonance stationary ILMs are observed below the bottom of a linear frequency band of the lattice. With decreasing driver frequency the width of the ILM changes in a step-wise manner as does the softening of the barrier between site-centered and bond-centered ILM locations in agreement with theoretical expectations. However, the steps show hysteresis between up and down frequency scans and such hysteresis inhibits the free motion of ILMs.
}

\begin{document}

\maketitle

\section{Introduction}
Very rare are lattice solitons, named by Toda when he discovered the integrable equations of motion for his monatomic lattice potential.\cite{1} Such lattice solitons travel freely but localized excitations in physical systems, characterized by nonintegrable lattices, loose energy during translation. Driven intrinsic localized modes (ILMs), that have been experimentally observed in a micromechanical cantilever array, clearly display pinning\cite{2,3}, and also show that a traveling ILM can only exist in a very small lattice requiring very special conditions.\cite{4,5} It is the barrier between site-centered and bond-centered lattice positions of the localized excitation that inhibits translation. Such a barrier is often associated with the Peierls-Nabarro (PN) potential in analogy with defect movement in a lattice.\cite{6} 

The topic of free-traveling versus site-pinning is of central interest, since ILMs are not restricted to motion in 1D so the control of the pinning effect for an ILM would open a new avenue. In this direction a saturable nonlinearity in a discrete nonlinear Schr\"{o}dinger (dNLS) equation has been proposed by Had\v zievski et al.\cite{7,8}.  They showed that the magnitude of the PN potential as a function of energy oscillates between positive and negative values and that this alternating sign changes the stability of an ILM between the two neighboring site centered and bond centered locations. Where the PN barrier is zero transition points exist and the ILM can move freely.   (Note that dNLS is nonintegrable and therefore it supports ILMs but not lattice solitons. These localized excitations can move through the lattice with a small, but non-zero deceleration\cite{9,10}. ) Interesting properties for such a saturable nonlinearity were proposed and many theoretical follow on publications have appeared.\cite{8,11,12,13,14,15,16,17,18,19,20}  

Nonlinear transmission lines have been considered as one of the convenient experimental tools with which to study excitations in 1-D nonlinear lattices.\cite{21,22,23,24,25,26} Recent experimental work has focused on ILMs\cite{27,28} where lattice discreteness plays an important role. Such studies have ranged from manipulation of a stationary ILM\cite{29}; to spatial control of slowly traveling ILMs\cite{30}; to generation of ILMs by sub-harmonic driving.\cite{31} For the earlier soliton experiments, reverse biased diodes have been used where the capacitance decreases with increasing pulse voltage (depletion capacitance). The lattice for such a case has an acoustic type dispersion curve, e.g., it starts from origin.\cite{Kuusela92,Remoissenet,24} (The Fourier transform components of the soliton include  zero frequency.) A soliton can form above the dispersion curve due to the decreasing capacitance with increasing pulse amplitude. On the other hand, an ILM is a vibrating localized mode somewhat like an envelope soliton. The capacitance of a diode increases with AC voltage amplitude because of injected carriers during the forward bias duration. This diffusion capacitance is much larger than the depletion capacitance. The dispersion curve for the ILM has a frequency gap and the ILM forms in the gap region because of the increasing capacitance with voltage amplitude.\cite{29}

By using a saturable electrical nonlinear lattice we show here through experiments and numerical simulations that softening of the barrier at stability transitions does exist. Our electrical lattice with on site nonlinearity can be converted to the dNLS\cite{32} and compared with the previous theoretical work. By following the frequency of a localized linear mode associated with the ILM the magnitude of the barrier is indirectly determined through measurement of the ILM lateral oscillations.\cite{2} 

This letter begins with the description of two experimental procedures: one associated with the production of an autoresonant  ILM in a nonlinear lattice and the second, with the linear response measurement of the ILM spectrum. They provide a way to explore the driver frequency dependence of the natural frequency (NF) of the ILM as well as the 1st-linear local mode (LLM)\cite{33} produced by the ILM.  The difference frequency between the NF mode and the driven ILM monitors the bifurcation point of the ILM while the difference between the 1st-LLM and the ILM measures the barrier height between site centered and bond centered locations. The observation of hysteresis in these latter transition point experiments and associated simulations was not predicted in the earlier theories. 

\section{Experiment}
\begin{figure}
\onefigure{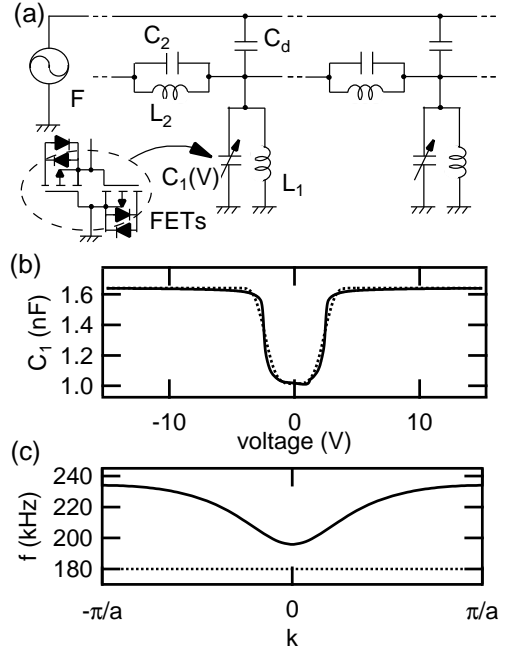}
\caption{(a) Electronic circuit lattice containing MOS nonlinear capacitors. Two FETs are connected anti-parallel, forming the nonlinear capacitor  $C_1$. One unit cell is made from a nonlinear resonator by  $C_1$ and  $L_1$, and a linear coupler  $C_2$ and $L_2$. Diodes near FETs are for discharging at drain electrodes, and reducing the hysteresis of the $C$ vs $V$ curve.  The lattice is excited uniformly by an oscillator at frequency $F$ through small capacitors  $C_d$. (b) Capacitance as a function of the applied DC voltage (solid) and a model curve used in simulations (dashed). Saturation of the capacitance provides the saturable nonlinearity. (c) Dispersion relation for the linear lattice circuit (solid curve). Dashed line indicates a typical frequency for ILM generation, below the bottom of the plane wave band.}
\label{fig.1}
\end{figure}

Figure~\ref{fig.1}(a) shows two unit cells of the 16 cell nonlinear electrical lattice. Each cell is composed of a nonlinear capacitor $C_1$  and a coil $L_1$=626 $\mu$H as a nonlinear resonator, and $C_2=660$ pF and $L_2$=626 $\mu$H  to provide linear coupling between adjacent resonators. The nonlinear capacitor consists of two anti-paralleled p-channel MOS-FETs (2SJ680). When the gate is positively biased, electrons in an N-semiconductor (for the p-channel FET) are accumulated below the gate between an oxide and semiconductor, and the boundary works as a conductive sheet. The capacitance between the gate and source electrodes is large with its value limited by the thickness of the oxide. When the gate is negatively biased, the semiconductor is inverted to P-type and the boundary layer forms a conductive sheet between drain and source electrodes. The capacitance between the drain and the gate is again large with its value limited by the thickness of the oxide. At a small bias voltage between accumulation and inversion, the capacitance is small because carriers at the boundary are depleted and the conductive sheet is not formed. Two anti-paralleled capacitance as a function of applied DC bias is shown in Fig.~\ref{fig.1}(b). One of the two diodes is a parasite diode of the FET, and the other one is an externally connected PN diode for discharge of stored charge to the drain electrode. Without it, the C-V curve in Fig.~\ref{fig.1}(b) produces a small hysteresis at the bottom of the curve and causes hysteresis damping. Since the capacitance increases with absolute voltage, the resonance frequency decreases with increasing amplitude (soft nonlinearity); however, the decrement saturates abruptly when the voltage becomes large. Figure~\ref{fig.1}(c) shows the linear band frequencies as a function of wave number. Because of the soft nonlinearity, an ILM can be generated below the bottom of the extended wave band. The dashed line indicates a typical driver frequency. The lattice is uniformly excited by an oscillator (10V) through a small capacitor  $C_d=34.3$pF. With this arrangement one can generate an autoresonant ILM where the driver frequency controls the amplitude of the ILM and hence its energy. The new feature produced by the saturation is that now the ILM width changes as well. 

\begin{figure}
\onefigure{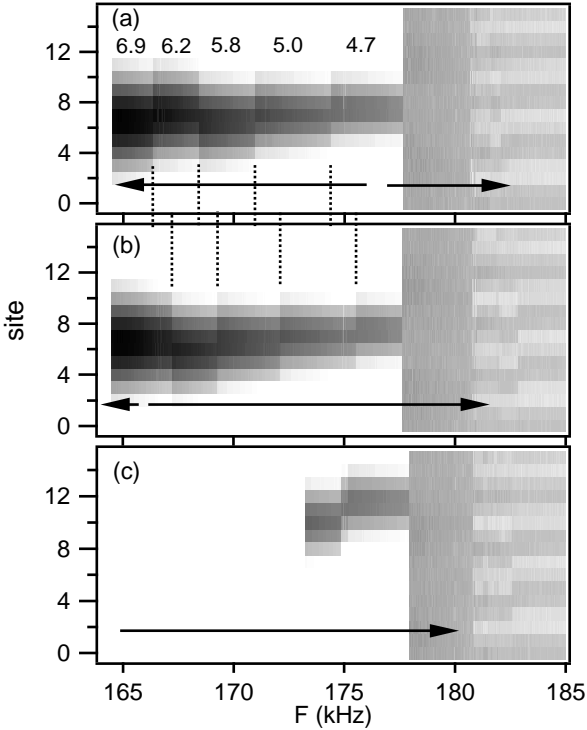}
\caption{Voltage amplitude pattern as a function of the driver frequency, for various scanning directions of frequency. Darker means larger oscillating amplitude. Arrows indicate frequency scanning directions. (a) The frequency was scanned down or up from 176 kHz. The initial ILM was seeded by using an impurity capacitor. The ILM width changes in a step wise fashion as the frequency is decreased. Numbers indicate full width at half maximum of the ILM. The shape changes alternately between bond-centered and site-centered across this range. At the higher frequency region, standing wave patterns appear. Their wavelengths change with the frequency. Vertical dashed lines indicate frequencies of step transitions. (b) Increasing the frequency from 165 kHz gives the same step changes. Note hysteresis between up and down scans. (c) Increasing the frequency from lower than stable ILM region. The ILM appears around 173.3 kHz, followed by the traveling wave packets, and standing wave patterns.}
\label{fig.2}
\end{figure}

Figure~\ref{fig.2} shows the modulus of the ILM voltage pattern as a function of the driver frequency. A darker image indicates a larger voltage. These patterns are measured from various starting conditions: (a) starting from the ILM at 176 kHz, (b) starting from the ILM at 165 kHz and (c) starting from no ILM state at 164 kHz. The patterns observed above 180.9 kHz are standing waves. The gray band from 177.6 to 180.9 kHz is due to traveling wave packets. The stationary ILM is generated either by simply lowering the driving frequency, or by seeding (using another capacitor as a temporary impurity). A stable ILM is observed between 177.6 kHz and 164.5 kHz as shown in Figs.~\ref{fig.2}(a) and (b). Below this lower value, the ILM disappears. Once that happens, the ILM cannot reappear till a frequency of 173.2 kHz is reached as shown in Fig.~\ref{fig.2}(c). Also an ILM can be generated if the frequency is decreased from that value. 

The observation of increased amplitude and step widening of the ILM in the stable frequency region shown in Figs.~\ref{fig.2}(a)-(b) is a feature of the saturable nonlinearity. In this figure, the ILM width nearly doubles as the frequency is decreased over the autoresonant range. In addition, the position repeatedly alternates between bond-centered and site-centered locations across the figure. Finally hysteresis in the frequency response of the ILM is observed when the frequency is tuned up or down as indicated by the vertical lines. Different transition frequencies occur for step-widening, see Fig.~\ref{fig.2}(a), and step narrowing, Fig.~\ref{fig.2}(b).

\begin{figure}
\onefigure{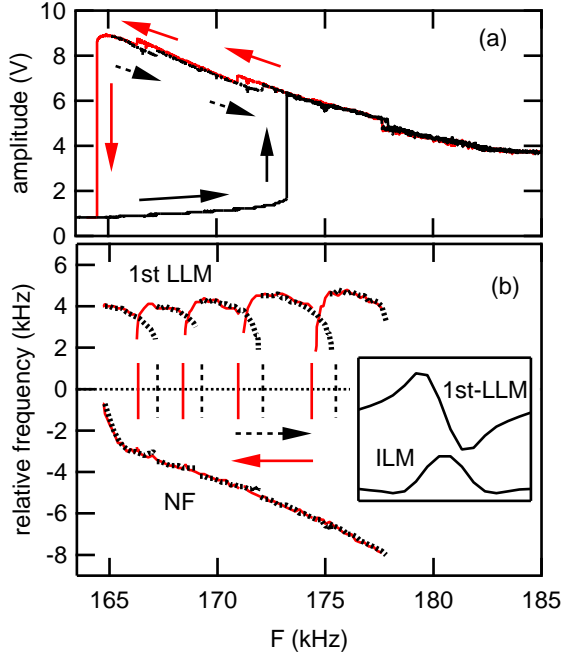}
\caption{(a) The maximum amplitude as a function of the driver frequency. Arrows indicate directions of frequency scanning. (b) Natural frequency (NF) of the ILM and the 1st-linear local mode (LLM) frequency as a function of the driver frequency. These modes are excited by a small amplitude probe driver and the perturbation is applied to one side of the ILM through a small capacitor (2pF). The small vibration of the 1st LLM is detected with a lock-in amplifier. Solid and dashed curves are for down-scanning and up scanning case, respectively. Vertical solid (dashed) lines at the center indicate frequencies of step transitions observed in Fig. 2 for down (up) scanning case. NF softens at 165 kHz identifying the approach to the bifurcation point for the ILM. The 1st-LLM frequency measures the pinning barrier of the ILM. Inset shows vibration shapes of the ILM and the 1st-LLM. The amplitude of the ILM is divided by 1000.}
\label{fig.3}
\end{figure}

Figure~\ref{fig.3}(a) presents the maximum amplitude of the ILM as a function of the driving frequency, $F$. Ignoring the small features for the moment the overall hysteresis loop looks very similar to that found for a Duffing oscillator with negative nonlinearity. Next linear response spectroscopy is used to measure the size of the barrier height between the site centered and bond centered locations as a function of the driving frequency. A sinusoidal perturbation is applied to one side of the ILM to excite its deformational vibrations. In this way we observe the frequency of the first linear local mode (1st-LLM).\cite{33,34}  Its amplitude is less than 1/1000 of the driver used to maintain the ILM so not to destroy the nonlinear state. This deformation is detected with a lock-in amplifier. By sweeping the sinusoidal perturbation frequency, we obtain a response spectrum showing the natural frequency (NF) of the ILM and up to several LLM resonances.  Such deformational vibrations appear as sidebands close to the ILM center frequency. The NF is easy to recognize since it occurs near by the ILM frequency, on the opposite side to the linear band, while LLMs are observed between the ILM and the extended wave band. Sometimes nonlinearly mixed NF and LLMs are observed on opposite sides, but these are smaller in peak heights. Among the several peaks, we focus on the 1st-LLM, because it corresponds to the lateral vibration of the ILM. This identification can be understood as follows. Vibration shapes of the ILM and the 1st-LLM at 177kHz are shown in the inset of Fig.~\ref{fig.3}(b). When the 1st-LLM is excited, they interfere with each other, and at a given instant the ILM has larger amplitude at one side but smaller amplitude at another side. Since the ILM and 1st LLM frequencies are different, this imbalance changes with time and the result is lateral vibration of the ILM at the difference frequency. The identification of the NF and 1st-LLM resonances in the spectra was made by checking the vibration spatial shape experimentally and by comparing these results with simulations. LLMs are located between the ILM and the band in frequency, so that they have positive difference frequency for the soft nonlinear case. On the other hand, the NF is on the opposite side of the LLM relative to the ILM, so that it is a negative difference frequency.  

Figure~\ref{fig.3}(b) presents the experimental findings for the frequency dependence of both linear modes as a function of the driver. The ordinate is the difference frequency with respect to the ILM.   The dots represent the results when $F$ increases and the solid lines, the results for the opposite case. The vertical lines identify the hysteresis effect in the transition point frequencies. As shown in Fig.~\ref{fig.3}(b), the NF difference frequency softens as it approaches the end of the stable region.\cite{34} This phenomenon is the same as the saddle-node bifurcation of the Duffing oscillator. The 1st LLM difference frequency softens as it approaches the step transitions from either side. It suddenly increases at the transition where the ILM width changes.  Once the transition takes place, one must reverse the frequency to recover the original shape of the ILM. The difference frequency continuously softens also in this case, and again jumps when the ILM width is reversed.  In other words, the barrier disappears when the step transition happens at a particular driver frequency, and the ILM shape changes from bond-centered to site-centered or vice-versa. However, once this happens, the barrier to change back to the other shape re-appears. To eliminate the barrier, one must tune the driver frequency again. The existence of hysteresis in the step transition frequencies indicates the possibility of different mechanisms for step narrowing and step widening transitions but the key point is that even at a transition a barrier remains.

\section{Comparison with numerical simulations}
We have performed numerical simulations using the equations of motions\cite{29} for the nonlinear electrical lattice to compare with experiment.  For the $n$th site the charge time dependence is given by

\begin{equation}
\label{eq.1}
\begin{array}{l}
 \frac{{d^2 Q(V_n (t))}}{{dt^2 }} = C_2 \frac{{d^2 }}{{dt^2 }}[V_{n - 1} (t) - 2V_n (t) + V_{n + 1} (t)] \\ 
  - C_d \frac{{d^2 }}{{dt^2 }}V_n (t) + \frac{1}{{L_2 }}[V_{n - 1} (t) - 2V_n (t) + V_{n + 1} (t)] \\ 
  - \frac{1}{{L_1 }}V_n (t) - \frac{1}{\tau }\frac{{dV_n (t)}}{{dt}} + C_d \frac{{d^2 }}{{dt^2 }}V_d (t) \\ 
 \end{array}
\end{equation}
where $Q(V_n(t))$  is the charge in  $C_1(V)$, $V_n(t)$ is the voltage across   $C_1(V)$,  $V_d$ is the voltage of the oscillator, and $\tau =41.7\mu $s is the damping time estimated by experiments. The nonlinear capacitance $C_1(V)$ is approximated by  $C_1 (V) = k_0  + k_1 \exp \left[ { - \left( {V/k_2 } \right)^4 } \right]$, where  $k_0=1.64$nF,  $k_1=-0.627$nF and $k_2=2.58$V. The model $C$ vs $V$ curve is shown by the dashed curve in Fig.~\ref{fig.1}(b). By using the relationship  $dQ=C_1dV$, the left hand side of Eq.~(\ref{eq.1}) becomes

\begin{equation}
\label{eq.2}
\frac{{d^2 Q(V_n (t))}}{{dt^2 }} = C_1 \frac{{d^2 V}}{{dt^2 }} + \frac{{dC}}{{dV}}\left( {\frac{{dV}}{{dt}}} \right)^2. 
\end{equation}
Simulations are then carried out using Eqs.~(\ref{eq.1}) and (\ref{eq.2}).  The 1st-LLM frequency and NF are obtained in a similar way as with experiments. A weak perturbation is applied to excite the linear local modes associated with the ILM, and their response is detected numerically. The two modes of interest are identified in the response spectrum by checking their eigenvector shapes. The details of the response calculation are given in Ref.\cite{34}.

\begin{figure}
\onefigure{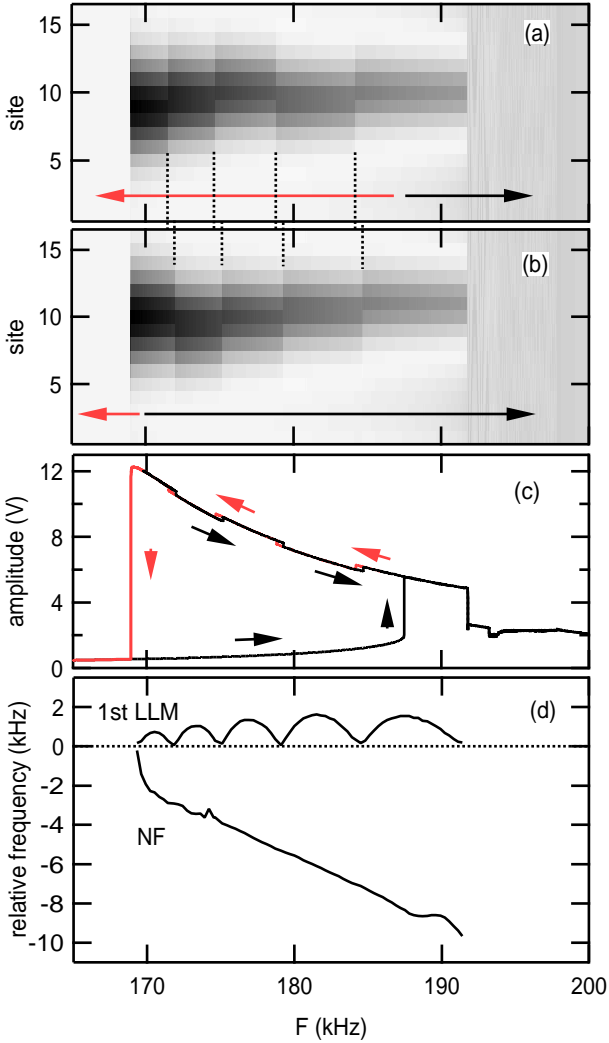}
\caption{Simulation results for the autoresonant ILM as a function of driving frequency, using Eq. (1) to describe the nonlinear lattice. (a) The ILM amplitude pattern as a function of the driver frequency. The frequency was scanned up or down starting from 186kHz. The ILM width increases in a stepwise manner for the down scanning case, while the traveling pattern appears at higher frequencies. Dashed vertical lines indicate transition points.  (b) For increasing frequency beginning at 169.8 kHz. Step narrowing is observed and the transition frequencies are slightly higher than for the down scanning case in (a). (c) The maximum amplitude as a function of the driver frequency. Hysteresis is found between up scanning from 170 kHz and down scanning from 190 kHz. (d) NF and 1st-LLM frequency. Softening of the NF mode as it approaches the fundamental bifurcation at the lower end of the stable ILM region. Also softening is observed for the 1st-LLM at step changes. }
\label{fig.4}
\end{figure}

Simulation results are summarized in Fig.~\ref{fig.4}. Figures~\ref{fig.4}(a)-(b) show the modulus of the ILM voltage pattern in the lattice as a function of the driving frequency. These results should be compared with Fig.~\ref{fig.2}(a, b). In Fig.~\ref{fig.4}(a), the simulation is started in the ILM state, and the frequency is either scanned up or down. As the frequency increases, the ILM disappears and propagating wave packets appear. These are followed by a chaotic excitation\cite{35}, then, another wave pattern is seen. If the frequency is decreased, the ILM broadens in a stepwise manner similar to the experiments. Once the frequency is below the stable region, the ILM disappears. In approaching this point from the stable ILM region, the NF softens as shown in Fig.~\ref{fig.4}(d). The sweep in the opposite direction is presented in Fig.~\ref{fig.4}(b). The vertical lines identify a smaller hysteresis effect than observed in experiment but it is still clearly evident. Hysteresis between the step widening and narrowing transitions is also evident from the amplitude plot in Fig.~\ref{fig.4}(c) when the driver frequency is scanned up or down in the stable ILM region. These results are similar to those shown in Fig.~\ref{fig.3}(a). Figure~\ref{fig.4}(d) demonstrates that the 1st-LLM difference frequency softens at both sides of a transition.   [The small hysteresis is not observable here.] Although not shown  the number of LLMs increases when step-widening is observed.

\section{Discussion}
From both experiments and simulations, the bifurcation on the low frequency side of the stable ILM region is the saddle-node bifurcation associated with softening of the NF of the ILM, similar to what was observed in the earlier micro-mechanical cantilever array experiments.\cite{34}  From Figs.~\ref{fig.3}-\ref{fig.4} the mechanism of the bifurcation at the step widening and narrowing transitions observed in Fig.~\ref{fig.2} is clearly due to a decrease in the barrier height between the neighboring lattice locations of the ILM.  The increase in the number of LLMs at the step widening transition occurs because of the larger frequency gap between the plane wave spectrum and the ILM. At the high frequency side of the stable ILM region, when it is close to the plane wave band, the 1st LLM softens towards that transition both in experiments and simulations, as shown in Figs.~\ref{fig.3}(b) and~\ref{fig.4}(d). In the mechanical cantilever array, the transition is somewhat different. The coalescence of the four-wave mixing partner of NF with the band mode is the signal of the bifurcation transition, which makes the ILM unstable. 

In comparing our results to the theoretical work by Had\v zievski et al.\cite{7} we see that in both cases, the ILM expands step-wisely as its amplitude increases, as shown in Fig. 2 of Ref.~\cite{7} and our Figs.~\ref{fig.2}-\ref{fig.4}. The width tends to widen by roughly one lattice site at each step, indicating alternating stability between site-centered and bond centered states. Between such states, the PN energy difference crosses the zero line in Fig. 1, Ref.~\cite{7}. The softening of the 1st-LLM frequency in Figs.~\ref{fig.3}(b) and \ref{fig.4}(d) at the transition corresponds to decreasing of the PN barrier height.  The zero-line is crossed as indicated by the stability change between the site-centered and bond-centered shapes at the transition locations shown in Fig.~\ref{fig.2}(a, b). A necessary difference between theory and experiment is that to maintain steady state in experiment a driver is required to counteract damping while in theory neither is considered. Yet the saturable nonlinearity produces the same step widen transitions with zero barrier height in both the circuit equation and nonlinear Schr\"{o}dinger equation cases. 

The physical requirement of a driver to counter damping leads to hysteresis in the transition points. Note that the experimentally observed hysteresis is larger than found in simulations, but still occurs in both cases. Since the simulation model ignores the possibility of nonlinear damping it represents the simplest possible driven model. This elemental property of hysteresis makes it unlikely that traveling ILMs can be generated in physical lattices.  Because the hysteresis produces zero barrier height for different frequency conditions for site-centered and bond-centered locations, both cannot be satisfied simultaneously. 

\section{Conclusion}
We have experimentally observed step widening/narrowing of the ILM in a saturable nonlinear lattice, caused by the interchange of stability between site-centered and bond-centered ILM locations. The 1st-LLM difference frequency softens as a step transition is approached. This signature indicates softening of the lateral barrier height at a stability transition consistent with previous theoretical work\cite{7}, except here we find hysteresis at each such transition, both in experiment and in simulations. The larger hysteresis, observed in experiments, may be caused by nonlinear damping.  Because of hysteresis the zero barrier condition for both site- and bond- centered locations cannot occur simultaneously and such is required for uninhibited ILM transmission in a physical lattice with saturable nonlinearity.

\acknowledgments
MS was supported by JSPS-Grant-in-Aid for Scientific Research No. 25400391. AJS was supported by Grant NSF-DMR-0906491.

\providecommand{\noopsort}[1]{}\providecommand{\singleletter}[1]{#1}%

\end{document}